\documentstyle[12pt]{article}
\nonumber

\begin{document}

\begin{center}
{\large\bf Particle propagation upstream of CIR shocks        }
\end{center}
{\large M. Savopulos}\\
 Information Dept, Education and Technological Institute of Thessaloniki, Greece\\
{\large J. J. Quenby and M.K. Joshi}\\
Physics Department, Imperial College, London, SW7 2BZ, UK\\
{\large M. Fraenz}\\
Queen Mary and Westfield College, Astronomy Unit, Mile End Rd., London E1 4NS, UK.\\
{\large\bf Abstract}
The first high solar latitude pass of the Ulysses spacecraft revealed the 
presence of MeV particle increases up to latitudes above 60 degrees, well 
outside the CIR belt but associated in time with the regular passage of 
these plasma interfaces at more equatorial latitudes. The particle 
increases have been explained variously as due to diffusion from a  
field line connection with a CIR at a greater distance, perpendicular 
diffusion in latitude, propagation from the inner Corona or an acceleration process on a connecting field line. Numerical solutions to the 1-D propagation 
equation upstream of a CIR shock for reasonable diffusion mean free 
paths are shown here to limit the source of the increases to within about 
2 AU of the CIR's, asssuming the CIR's either trap or accelerate energetic 
particles and they diffuse from the nearby CIR interface. 
The problem of allowing propagation from the CIR's is eased if the CIRs are
located according to the predictions for the current sheet with a solar source
surface at $2.5 R_{s}$, rather than  at $3.25 R_{s}$. Energetic electron observations showing delays with respect to the CIR-associated ions suggest possibly an acceleration in the inner corona, rather than in the CIR-associated shock. A more likely explanation, however, is the ability of electrons to take a longer a longer route, from beyond the point of observation, than for that of the ions and yet arrive within a few days of CIR closest approach time.\\
1. {\large\bf Introduction}\\  
In the course of the first Ulysses southern latitude polar 
pass it was found that whereas the interaction regions between
high speed and low speed solar wind driving CIR's were limited to solar latitudes between
$13^{\circ}$ and $40^{\circ}$, associated energetic particle increases appeared up to $64^{\circ}$ in the case 
of 0.5-1.0 MeV protons and to $75^{\circ}$ in the case of 50 keV electrons. Despite the well known longitudinal
spreading of such particle increases in equatorial regions (O'Gallagher and Simpson, 1966  ), this observation 
constituted a 
major surprise to those working on interplanetary energetic particle transport. It clearly provides
crucial information on the relative importance of various competing plasma and energetic particle
 transport processes (Fisk, 1997). The particles may
have originated in the CIR's due to interplanetary diffusive shock acceleration (Palmer and Gosling, 1978),
or they may have been mainly accelerated close to the sun and carried out in a trapping region 
(Lim et al., 1996). In either case, cross-field diffusion from the CIR to Ulysses (Kota and Jokipii
, 1995) or sunward diffusion along a field line connecting to an expanded CIR beyond the Ulysses orbit
(Keppler et al., 1995, Lanzerotti et al., 1995) or sun-ward diffusion along a field line constrained to move in
latitude under the combined influence of differential solar rotation and differing wind speeds which eventually
meet the streamer belt ( Fisk, 1996) are all possible modes of propagation for the high latitude energetic
 particles. A third alternative is that particles, originally accelerated at the sun, diffuse in the inner 
corona before they move out along connecting field lines to Ulysses ( Quenby et al., 1996). In the following we
describe data relevant to the high latitude increases and discuss possible models.\\
An additional complication in the observational data lies in the delayed onset of the CIR-related energetic
electron increases, relative to ions, above the low latitude streamer belt, Simnett and Roelof (1995). \\
2. {\bf Data}\\
The helium data 
to be employed has already been presented by Keppler et al., (1995) and was obtained from the Ulysses
EPAC experiment. 4 identical, three-element semiconductor telescopes mounted at different angles to the 
spin axis yielded 8 sector information over $80 \%$ of the solid angle The geometry factor to measure 0.3 to 1.5
MeV protons and 0.4 to 6 MeV/N heavy ions was 0.08 cm$^{2}$ster. \\
We confine ourselves to the use of 0.4-1.0 MeV/N omnidirectional helium data obtained on the first Ulysses 
southern solar pass. Figure 1 is reproduced from Keppler et al., (1995) in order to clarify subsequent
 discussion.
The regular series of energetic alpha particle increases, 26 days apart are shown plotted against time,
 spacecraft-sun distance and heliolatitude.
In general, the greater part of these particle increases lie between well-identified forward-reverse shock pairs
 in the low latitude 'streamer belt' 
but cease to be enclosed by significant plasma discontinuities at 
high latitudes (Phillips et al, 1995). $40^{\circ}$ is taken as the
transition point between 'low' and 'high' in our subsequent analysis. 
  A mean spectrum for energetic He ions within 
a CIR is established from data obtained
close to the reverse shocks identified between latitudes -32$^{\circ}$ and-38$^{\circ}$. We will also use data
at higher latitudes, again coinciding with the regular particle increases which seem CIR-associated, at the 
distance (AU) and southern latitude (degrees) pairs; (4.2,41),(4.0,44),(3.8,47),(3.6,52),(3.4,56),(3.2,60).\\
Electron and ion data are obtained from the Ulysses HI-Scale experiment, (Lanzerotti et al., 1992). We use
the archived data from the thin, single solid-state detector system with active anti-coincidence and thin foil 
separation of electrons in the energy range 30 keV to 300 keV and ions in the energy range 300 keV to 5 MeV.
The telescopes involved have solid angles of 0.48 cm$^{-2}$ sr.\\
3.{\bf Shock Expansion Model}\\  
Since it will turn out that the models for the high latitude increases 
most stringently constrained by the above data are those involving su-ward diffusion
from a CIR source located well beyond Ulysses, we will centre our numerical analysis on one of these, namely
 the expanded CIR beyond the spacecraft orbit.
In order to determine the likely range over which diffusion takes place, it is thus necessary to establish 
plausible estimates
of the expansion history of a CIR as it moves beyond Ulysses. Gosling et al., (1993) pointed out that 
the reverse
shock moved poleward. Figure 2 is a cartoon representing a meridian plane cross-section of an interaction 
region in the southern hemisphere with the 
equator-ward forward shock expansion and the pol-ward reverse shock expansion. Ulysses is located at a
 latitude such that
it does not directly see the CIR interaction region but lies on a radial line from the sun which subsequently 
intersects the expanded CIR. A simple estimate of the expansion is obtained following Quenby et al., (1995), 
who assumed a 
radial speed of 500 km/s and found the pole-ward or latitude 
expansion speed arising from the excess pressure within the CIR to be the 
 Alfven speed of 150 km/s. In this calculation, it is further assumed that the 
extent of the streamer belt remains constant in
it's upper latitude boundary at a given radial distance throughout the southern latitude pass. Starting from the observed       
extent of the streamer belt as reaching 36$^{\circ}$S at 4.4 AU, the CIR is then found to lie on the same 
radial vector as Ulysses at the following pairs of distances and latitudes; (5.8,41),(6.9,44),(8.2,47),(10.8,52),
(13.6,56),(17.1,60). This calculation of the CIR expansion is to be regarded as the maximum possible and 
indeed, the simulations by Pizzo, (1991)  suggest a reduction
in the rate of latitude increase beyond $\sim$ 10 AU. \\ 
It is not easily established how the intensity increase inside the CIR varies with position. We have tried
correlations with radial distance, Ulysses latitude and the maximum extent of the neutral sheet at the time.
There is only a poor correlation with the later two parameters but while the radial increase yields the
dominant correlation, we cannot from our data and epoch extrapolate beyond 5 AU and  
so we take a favourable case and assume that the intensity increase stays constant. Thus we take the mean intensity
 found previously as applying both as the source intensity inside the CIR
for the (4.2,41) observation and to the subsequent points where Ulysses is connected on a
 solar radius vector to the CIR
which has propagated beyond and above this position. Infact, if the Van Hollebeke et al., (1978) data on 
 the equatorial plane radial
dependence of the peak CIR particle intensity applies, the source flux within the CIR will diminish
beyond 4-5 AU.   All the intensities corresponding to the 
above pairs of positions are assumed to refer to observations upstream (or sun-ward) of the expanded reverse 
shock of the CIR. These intensities, normalised to the intensity inside the CIR (shock and observation 
at 5.8 AU)    
are plotted in figures 3 and 4, together with the estimated upstream distances on the Quenby et al., (1996) model.
\\ 
If the connection between the CIR from beyond Ulysses to the spacecraft is via a field line moving in latitude
according to the Fisk, (1996) model, the radial distance to be moved is typically 10 AU for a line seen at 
70$^{\circ}$ in the inner heliosphere. Thus calculations on the flux expected upstream of shock on the 
expanded CIR model are applicable to geometries based upon the Fisk, (1996) model
provided in each case, the dominant mode of particle diffusion is radial.\\
4.{\bf Semi-quantitative consideration of the possible propagation}\\ 
The ability of MeV particles to propagate $\sim 10$ AU to be seen with the observed 
intensity and to be reasonably in phase with the established low latitude solar rotation periodicity are 
critically dependent on the radial or perpendicular diffusion mean free path adopted. It is difficult to better
the 'Palmer Consensus' value of 0.1 AU (Palmer, 1982) as applying in a wide rigidity and distance range 
for $\lambda_{rr}$. However a detailed study of the time variability in the magnetic fluctuation spectum
on an hour to hour basis and of the anisotropy injection profiles of long lived particle events by 
Wanner and Wibberenz, (1991) found $\lambda_{\|}$ varying between 0.01 and 1.0 AU. Reames (1999) maintains
that both prompt and gradual SEP events are consistent with $\lambda_{\|} \sim$ 1 AU.   
The 'Palmer' value is in accord with recent realistic, magnetometer based computations of the 
parallel mean free path
(Drolias et al., 1997) over the Ulysses orbit which yield $\lambda_{\|}\sim 0.1$ AU provided we are at $\sim 1$ AU.
 There is, however,
already a puzzle beyond $\sim 5$ AU in adopting this relatively low value of $\lambda_{\|}$
where $\lambda_{rr}=\lambda_{\|}cos^{2}\chi$ where $\chi \sim 70^{\circ}$
or $\lambda_{rr} \sim 0.01$ AU , unless $\lambda_{\perp} \sim \lambda_{\|}$. \\
Using the expression for the diffusion time, $\tau$, to reach peak flux over distance, L, with velocity, v,
$\tau \sim (3/4)L^{2}/\lambda_{rr}v$, we find it takes 94 days to diffuse 10 AU, 23 days to diffuse 5 AU and
 90 hours to diffuse 2 AU. These numbers suggest that unless the 'source' of the high latitude increases is
within a few AU of Ulysses, the 27-day periodicity in the enhancements will be difficult to explain.\\
5.{\bf Quantitative radial back diffusion model}\\
In this section, we numerically solve the Fokker-Planck transport equation for the case of spherically
symmetric propagation back towards the sun from the reverse shock of a CIR. The Fokker-Planck with only
radial coordinate dependence and assuming a steady state is
\begin{equation}
\frac{1}{r^{2}}\frac{\partial}{\partial r}(r^{2}k\frac{\partial f}{\partial r})-(V-V_{s})
\frac{\partial f}{\partial r}+\frac{2}{3}\frac{Vv}{r}\frac{\partial f}{\partial v}=0
\end{equation}
Here k is the radial diffusion coefficient where $k=\lambda_{rr} v/3$
, f the distribution function, V the wind speed, $V_{s}$ the 
shock speed, r the solar radial distance and all quantities measured in a solar reference frame.
The equation expresses the facts that while the diffusive and adiabatic deceleration divergence terms depend 
on distance from the Sun, the convective motion must be considered relative to the shock. The boundary 
condition derived from continuity of streaming at the CIR reverse shock is
\begin{equation}
-k(\frac{\partial f}{\partial r})_{1}+\frac{v}{3}(\frac{\partial f}{\partial v})_{1}(V_{s}-V_{1})
=\frac{v}{3}(\frac{\partial f}{\partial v})_{2}(V_{s}-V_{2})
\end{equation}
where 1 and 2 refer to upstream and downstream conditions. Using the inverse compression ratio $\beta$,
$V_{s}-V_{2}=\beta(V_{s}-V_{1})$. The boundary condition equates upstream diffusive and convective flows with 
downstream convection with negligible diffusion. Following Savopulos et al., (1995), equations 1 and 2 are 
combined and numerically solved by inverting a quindiagonal matrix allowing $k=k_{\circ}v^{\alpha}r^{\epsilon}$
. $k_{\circ}$ will be quoted with v measured in units of $10^{8}$ m s$^{-1}$ and r in units of 
$10^{11}$ m.\\       
We show in figures 3 and 4 as continuous lines solutions to equations 1 and 2 for two relations for k. In 
figure 3, $\alpha=\epsilon=1$ and $\lambda_{rr}$=0.008AU at 1 MeV/N and 1AU. In figure 4, $\alpha=1.6,
\epsilon=0.0$ and the $k_{\circ}$ value of 5.1435 in the above units corresponds to $\lambda_{rr}=0.44$ AU
at 1MeV/N independent of distance. The top curve is the fit to the He spectrum inside the CIR while the
others correspond to the computed upstream spectra for the assumed shock-Ulysses distances on the model of
section 3 for the latitudinal spread of a CIR. These distances can similarly be employed in the model of the
latitudinal wandering of an interplanetary field line. The figures show that a variety of mean free path models
can fit the data (as demonstrated by Savopulos and Quenby, 1998) although independent propagation evidence 
clearly favours an intermediate mean free path value. However, both figures confirm that there is 
little prospect of fitting the intensity fall-off apparently observed upstream of the reverse shock except
within about 1 or 2 AU of the CIR, if either back diffusion model is adopted. \\
The fall in intensity due to convective sweeping and adiabatic expansion over $\sim$
 5AU also calls into question models 
requiring coronal plus interplanetary propagation.\\
Some significant reduction in the requirement on the distance travelled in latitude is gained by following 
Sanderson et al.,(1999) who employ the Wilcox Solar Observatory curent sheet locations based on the $2.5 R_{s}$
source surface model, rather than that with the source surface at
 $3.25 R_{s}$, which is often used. Both models assume potential 
fields and a radial source surface field, but the former adds a sharply peaked polar field to the surface
line-of-sight data while the latter assumes a purely radial surface field. Both models are reasonably,
but not entirely successful in predicting to the IMF (eg Hoeksema,1995). Using the $2.5 R_{s}$ model 
the curent sheet is found to have passed within about $5^{\circ}$ of Ulysses up to about day 60, 1994,
although there still remains a gap opening to $\sim 30^{\circ}$ at the last ion increase seen around day 160, 1994,
so the problem is only possibly partly solved with the model with the lower source surface. The active interaction region
thought to be operative at these times is not considered to extend much south of the maximum southward extent
of the neutral sheet at the Ulysses radius.   \\  
6. {\bf Interpretation Of 'Delayed Electrons'}
Figures 5a and 5b illustrate the delayed electron arrival relative to the ions above the streamer belt. 
Figure 5a runs from day 183 to day 273, 1993 and shows the electron channels 30-50 keV(E1), 50-90 keV(E2),
90-165 keV(E3), 165-300 keV(E4) and the ion channels 300-550keV (FP4), 550-keV-1MeV (FP6) and 1-5MeV (FP6).
The electron and ion increases up to day 225, corresponding to a South latitude of about $37^{\circ}$,
tend to peak at the same time, although the electron increase often extends to later times or further upstream
from the reverse shock, behaviour confirmed by similar plots at lower latitudes. By day 261,
 corresponding to a latitude of 
about $41^{\circ}$, the electron increase is delayed 2 days in the lowest energy channel at least.
 Figure 5b for the time
period day 274 to 365, 1993 shows a delay of about 2 days developing at all energies shown,  
again a number confirmed
by early 1994 data.  \\
In seeking to understand the electron data delay we note the conclusion of Palmer (1982)
 that the parallel electron mean free path
between 10 keV and 10 MeV lies in the interval 0.1-1 AU with a likely value $\sim 0.1$U AU. 
Scatter-free events with $\lambda \sim 1$ AU are judged to be relatively rare.
Assuming perpendicular diffusion over the shortest possible mean free path, let us attempt an interpretation of  
the extra electron time delay around day 343, 1994, when Ulysses was at $\sim 3.9$ AU at latitude 
$\sim 46^{\circ}$ south. According to Sanderson et al,(1999), the current sheet was at $\sim 40^{\circ}$
when it passed closest to Ulysses. Hence the distance over which perpendicular diffusion is to take place
from the CIR to Ulysses is $\sim 0.4$ AU. 
A way of estimating the MeV proton propagation time is to take a $\lambda_{\perp}\sim0.1 \lambda_{\|}$.
This is within an acceptable range and is favourable to explaining the electron data by the same propagation
mode. With$\lambda_{\|}\sim0.1$AU, $\lambda_{\perp}$=0.01AU. This yields a 1.7 day delay. 
Suppose a total diffusion time across the field above the CIR of 1.7 day  
(suggested by the proton diffusion data) plus the 2 days extra electron delay at 230 keV.
Then we require an electron $\lambda_{\perp}=0.0003-0.0004$ AU in the range 40-230 keV. 
There is thus no consistency with the
'Palmer consensus' value of the cross field $k_{\perp}/\beta \sim 10^{21}cm^{2}s^{-1}$ or $\lambda_{\perp}\simeq 7.10^{-3}$AU. 
Neither is the theoretical field line wandering estimate (Forman et al.,1974) of $k_{\perp}/\beta\simeq
4\times10^{20} cm^{2}s^{-1}$ satisfied, especially the velocity independence of this quantity. \\
Computations by Kota and Jokippi, (1998), suggest that the 100keV electrons, in contrast to the MeV ions,
can infact arrive from a CIR position further out than Ulysses. These authors employ a tilted dipole model
of the heliospheric field, with purely radial flow, that incorporates the forward and reverse shocks at the
interfaces between the slow and fast streams. The 3-D test particle transport equation is solved, including
already accelerated electron and ion spectra injected at the shock fronts. The simulations reveal a broader
time peak in the high latitude arrival with several days delay with respect to the ions, suggesting 
propagation predominantly from further out. \\
Using our 'semi-quantitative approach' as before, we estimate a three day delay for 100 keV electrons with
$\lambda_{\|}$ =0.1 AU would enable a distance of 1.8 AU to be traversed along a field line whereas this 
distance increases to 5.7 AU if $\lambda_{\|}$ =1.0 AU. Since these relativistic electrons are in a propagation
parameter parameter regime where diffusion dominates convection in the transport equation (1), the Fisk and
Lee (1980) approximation to the remaining, diffusion and adiabatic deceleration terms should yield the radial
intensity dependence upstream of the shock. This solution gives $f=F(v)(r/r_{s})^{2\beta/(1-\beta)}$
provided $ k=k_{\circ}vr$ with $r_{\circ}$ as shock location. Hence with $\beta=1/4, f \propto r^{2/3}$
and therefore the intensity reduction over 2-6 AU is not drastic. This fact can also be inferred from fig 4
which in the high velocity regime and for $\lambda_{\|}$ =0.44 AU corresponds to the diffusion dominated solution
regime. Note that the theoretical curves approach each other at the highest velocities, denoting a low spatial
gradient in the parameter space where the radial dependence is expected to be particle velocity and species
independent according to the Fisk and Lee (1980)  approximation. Clearly with an appropriate choice of the 
$k_{\perp}/k_{\|}$ ratio, one can arrange for the cross field diffusion electron intensity to be less
than the along the field diffusion electron intensity. Note this 1-D discussion of an essentially 3D
situation depends on $k_{\perp}<<k_{\|}$.\\       
An alternative approach to the electron delay is to assume most acceleration takes place close to the sun
within the streamer belt confines or with particle release mainly situated within this region. It is unlikely
that electrons appear at higher latitudes relative to protons because of a basic asymmetry in the output of
reconnection event aceleration because such events are orientated at random. In the IMF at the epoch of
observation, the gradient and curvature drifts will produce a drift velocity in a direction such that the
electrons appear later, but the drift speed is only is only $\sim 3\times10^{5}$ cm/s. Hence this is unlikely
to be a cause of the measured delay as insufficient latitude separation occurs. A possibility is that at
$\sim$ 100 keV or $\sim$ 0.3MV, $\lambda_{\|} \sim$1 AU. Then electrons accelerated within a few $R_{s}$
propagate easily to Ulysses, outside the CIR belt, after coronal diffusion while protons $\sim$0.4 MeV
or $\sim$30 MV are much attenuated in intensity in the coronal diffusion region or even in the IMF. This idea
presupposes that favourable acceleration conditions involving reconnection or shock propagation across closed
magnetic loops occur mainly within the streamer belt. The observation by Van Hollebeke et al, (1978), that 
within CIR's the proton intensity increases more than 100\% from 0.4 AU to 1 AU simply states that some interplanetary
acceleration occurs within the CIR but does not invalidate the idea of an important coronal component to acceleration.
A coronal diffusion model explains the reduction in
appearence rate with increased latitude for electrons noticed by Lanzerotti et al, (1995). CIR particle abundances
are mainly similar to solar wind abundances (Keppler, 1998), rather than flare abundances. However, the likely source
of the solar wind is small-scale reconnection events, so there can be no basic incompatibility in the source
abundances. Large-scale flares are clearly potentially different in plasma constitution. \\
7.{\bf Conclusions}\\
With accepted values of the radial mean free path in the few AU, few MeV/N region, the models for high solar
latitude CIR associated particle increases which involve diffusion over 5-10 AU, either from behind Ulysses or
directly from the Sun, have difficulty in fitting the relatively high intensities seen up to 60$^{\circ}$ when
spherically symmetric solutions of the Fokker-Planck equation are employed. The problem is enhanced by the fact
that favourable assumptions were made for the intensity of the source particles within CIR's beyond Ulysses 
and for the height of the CIR expansion in latitude.
There is an additional problem
in the long time required for such diffusion to take place rendering the observed phase coherence of the 
particle increases difficult to understand.
An alternative is to consider a perpendicular diffusion mean free path not much smaller than the parallel
$\lambda$, allowing the CIR associated fluxes to propagate only 1-2 AU to locations
directly above the shock in latitude. Use of the neutral sheet model with solar source at $2.5 R_{s}$ 
reduces the amount of latitude propagtion required although it remains significant.\\
The relatively large value of the perpendicular diffusion coefficient required in the above analysis fits with
the relatively small latitudinal gradients in low energy cosmic ray intensity seen by Ulysses which appear
significantly less than expected on full drift models with $K_{\perp}=0.05K_{\|}$ (see Drolias et al., 1997
and references therein).\\
100 keV electron increases observed above the streamer belt are not easily explained on the purely perpendicular
diffusion transport model with reasonable parameters. Instead, it is possible that a coronal source is the basic
accelerator and the scattering mean free path outside CIR's at sub MV rigidities is of the order of 1 AU. Most 
ion transport would then be effected by trapping within the CIR's while most electron transport would be first
via the corona and then in the quiet IMF. However, back diffusion from the expanding CIR beyond Ulysses is a likely
explanation for the delayed and extended electron arrival as fewer special assumptions arise in this 
last model.\\     
8.{\bf Acknowledgements}\\  
Helpful discusions with Bern Blake, Erhard Keppler, Bob Forsyth, Trevor Sanderson 
and Bernd Heber are acknowledged. PPARC visitors' support for M. Savopulos was gratefully received.\\
9.{\bf References}\\
Drolias,B.,Garaud,P.,Quenby,J.J. and Smith,E.J.:1997, Proc.25th. ICRC,1,229.\\
Fisk,L.A.:1996, JGR,101,A7,15,547.\\
Fisk,L.A.:1997, Proc.25th. ICRC,8,27.\\
Fisk,L.A and Lee,M.A.:1980, Astrophys. J. 237,620.\\
Forman,M.A.,Jokippi,J.R. and Owens.:A.J.,1974, Astrophys.J.,192,535.\\
Gosling,J.T.,Baker,D.N.,Bame,S.J.,Feldman,W.C.,Zwickel.R.D.and Smith,E.J.:1993, Geophys. Res. Lett.,20, 2789.\\
Hoeksema, J.T.: 1995, Space Sci., Rev.,72/1-2,137.\\ 
Keppler,E.,Franz,M.,Korth,A.,Reuss, M.K.,Blake,J.B.,Seidel,R,Quenby,J.J. and\\ 
Witte,M.:1995,Science,268,1013.\\
Keppler,E.:1998, Surveys in Geophysics, 19,(3 SIS),211.\\
Kota,J. and Jokipii,J.R.:Science, 268(5213), 1024.\\  
Kota,J. and Jokipii,J.R.:1998, Space Sci.Rev.,83,137.\\  
Lanzerotti,L.J.,Gold,R.E.,Anderson,K.A.,Armstrong,T.P.,Lin,R.P.,Krimigis,S.M.,Pick,M,.
Roelof,E.C.,Sarris,E.T.,
Simnett,G.M. and Frain.,W.E.:1992, Astron. Astro. Phys. Suppl. Series,92,349.\\ 
Lanzerotti,L.J.,Armstrong,T.P.,Gold,R.E.,Maclennan,C.E.,Roelof,E.C.,Simnett,G.M.,\\Thomson,D.J.,Anderson,K.A.,
Hawkins III,S.E.,Krimigis,S.M.,Lin,R.P.,Pick,M.,Sarris,E.T. and\\ Tappin,S.J.:1995,Science,268,1010.\\
Lim,T.L,Quenby,J.J.,Reuss,M.K.,Keppler,E.,Kunow,H.,Heber,B. and Forsyth,R.J.:1996,
Ann. Geophys.-Atmos. Hydrospheres Space Sci. 14(4),400.\\
O'Gallagher J.T. and Simpson, J.A.: 1966, Phys. Rev. Lett., 16, 1212,\\ 
Palmer,I.D. and Gosling,J.T.:1978,JGR, 83,2037.\\
Palmer,I.D.:1982, Revs. Geophys. Space Physics, 20, 335\\
Phillips,J.L., Bame,S.J.,Feldman,W.C.,Goldstein,B.E.,Gosling,J.T.,Hammond,\\C.M.,McComas,D.J.,Neubauer,M.,
Scime,E.E.,and Suess,S.T.,1995.: Science, 268, 1030.\\
Pizzo,V.J.,1991.: J. Geophys. Res., 96, 5405.\\ 
Quenby,J.J.,Drolias,B.,Keppler,E.,et al.: 1995, Geophys. Res. Lett.,22,3345.\\
Quenby,J.J.,Witcombe,A.,Drolias,B.,Fraenz,M. and Keppler,E.:1996,Astron. Astrophys. 316,506.\\
Reames,D.V.:1999, Space Sci. Rev.,90,413.\\
Sanderson,T.R.,Lario,D.,Maksimovic,M.,Marsden,R.G.,Tranquille,C.,Balogh,A.,Forsyth,R.J. and Goldstein,B.E.:
 1999, Geophys. Res. Lett.,26,1785.\\ 
Savopulos,M.,Quenby,J.J. and Bell,A.R.:1995, Solar Phys.,157,349.\\
Savopulos,M. and Quenby,J.J.:1998, Solar Phys.,180,479\\
Simnett, G.M. and Roelof,E.C.:1995,  Space Science Reviews,72,308\\  
Van Hollebeke, M,A,I., McDonald,F.B.,Trainor,J.H. and Von Rosenvinge,T.T.: 1978, JGR, 83, 4723,\\ 
Wanner,W. and Wibberenzz,G.: 1991, 22nd Int. Cosmic Ray Conf., Dublin, 3,221.\\  
{\bf Figure Captions}\\ 
Figure 1 Overview of the first Ulysses south high latitude pass showing the solar wind, magnetic field
strength and energetic helium flux as a function of heliolatitude.\\
Figure 2 Representation of a heliospheric meridian plane showing the oscillating neutral current sheet,
the developing forward and reverse shocks and the radial line through the Ulysses spacecraft.\\
Figure 3 Observed high latitude He intensities as a function of particle velocity measured in MeV/N units.
Positions of observation and modelled positions where the expanding CIR's reach the radial line from the 
sun through Ulysses beyond the spacecraft are indicated. Computational 
solutions corresponding to these upstream of the shock positions are shown as solid lines. The diffusion parameters
employed are stated at the bottom.\\
Figure 4 Similar to figure 3 for alternative diffusion parameters.\\
Figure 5 Ulysses electron CIR observations above the streamer belt, 1993, days 183-273.  Comparison
ion channels are also shown.\\ 
Figure 6 Ulysses electron CIR observations above the streamer belt 1993, days 274 to 365. Comparison ion channels
 are also show.\\
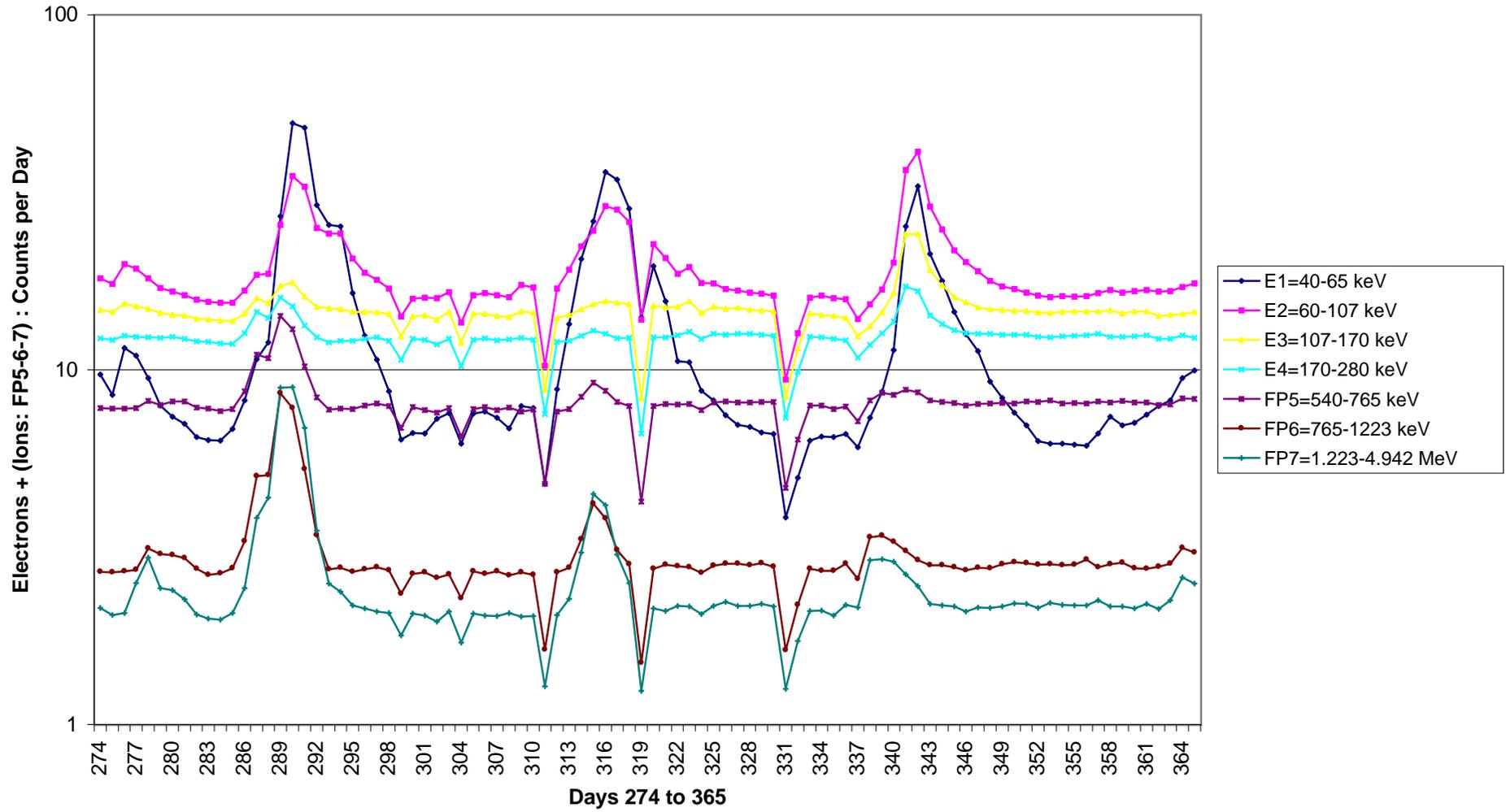
\end{document}